\documentclass[conference]{IEEEtran}
\IEEEoverridecommandlockouts
\usepackage{cite}
\usepackage{amsmath,amssymb,amsfonts}
\usepackage{algorithmic}
\usepackage{graphicx}
\usepackage{textcomp}
\usepackage{xcolor}
\def\BibTeX{{\rm B\kern-.05em{\sc i\kern-.025em b}\kern-.08em
    T\kern-.1667em\lower.7ex\hbox{E}\kern-.125emX}}
\begin{document}

\title{Capturing Stability of Information Needs in Digital Libraries}

\author{\IEEEauthorblockN{Christin Katharina Kreutz\IEEEauthorrefmark{1}, Philipp Schaer\IEEEauthorrefmark{1}, Ralf Schenkel\IEEEauthorrefmark{2}}
\IEEEauthorblockA{\IEEEauthorrefmark{1}\textit{TH Köln - University of Applied Sciences}, Cologne, Germany \\
\IEEEauthorrefmark{2}\textit{Trier University}, Trier, Germany \\
\IEEEauthorrefmark{1}\{christin.kreutz, philipp.schaer\}@th-koeln.de, \IEEEauthorrefmark{2}schenkel@uni-trier.de\\
}
}

\maketitle

\begin{abstract}
Scientific digital libraries provide users access to large amounts of data to satisfy their diverse information needs. Factors influencing users' decisions on the relevancy of a publication or a person are individual and usually only visible through posed queries or clicked information. However, the actual formulation or consideration of information requirements begins earlier in users' exploration processes. Hence, we propose capturing the (in)stability of factors supporting these relevancy decisions through users' different levels of manifestation.
\end{abstract}


\section{Introduction}

Bibliographic digital libraries (DLs) such as the ACM DL, Bibsonomy~\cite{DBLP:series/xmedia/HothoJBGKSS09}, or dblp~\cite{DBLP:journals/pvldb/Ley09} 
provide a wide variety of information to their users.
Within these systems, users can decide on the relevancy of information objects and satisfy their information needs, e.g., if a scientific publication is relevant to a topic. 
Users consider a multiplicity of relevancy indicators; they determine the relevancy of a document for a task on more than mere topical fit~\cite{Cool1993CharacteristicsOT}. Generally, which indicator is considered relevant partially depends on the application domain~\cite{DBLP:journals/ipm/BarryS98}.

Many works~\cite{kuhlthau2004seeking,Bates1989TheDO,Taylor1968QuestionNegotiationAI,DBLP:conf/jcdl/LiuS22,DBLP:conf/jcdl/HoeberS22} explain or model users' overall information seeking strategies.
Factors that users consider in their relevancy judgements on information objects change as their cognitive state changes in their information gathering process~\cite{DBLP:journals/ipm/Taylor12}.
However, research on the satisfaction of information needs focuses on general strategies of users~\cite{DBLP:conf/jcdl/HoeberS22,DBLP:conf/chiir/HoeberPS19} and less on changes in factors between different representations of information needs. 

In this work, we suggest investigating the persistence or change of users' considered factors throughout different manifestations in their information seeking strategies. 
The construction of information exploration and retrieval systems can be improved by analysing how users describe their information needs to humans, not only their formalised queries~\cite{DBLP:conf/chiir/ArguelloFFMZ021}. 
Additionally, Ingwersen~\cite{ingwersen} assumes that the change between a verbalisation and the actual conduction of a task leads to users compromising their information needs. 
In general, the longitudinal stability of users' defining factors of information needs is under-researched.

Therefore, we propose extracting and observing key factors from users' (1) general definition of an information need, (2) in an idealised retrieval process, (3) in the actual task conducted with an information system and over time (4).

\section{Related Work}


%

Research on users' information seeking behaviour and expression of information needs has a rich history:
Taylor~\cite{Taylor1968QuestionNegotiationAI} proposed a four-level continuum to describe the expression of information needs in the context of a person coming up with the formulation and satisfaction of their information need: $Q_1$ describes the actual visceral and linguistically inexpressible need for some type of information, $Q_2$ describes the conscious mental description for some type of information, $Q_3$ describes the verbalised need for some type of information, and $Q_4$ describes the compromised interaction with an information system to satisfy the need for some interaction.

Belkin et al.~\cite{DBLP:journals/jd/BelkinOB82a} were among the first to define information needs as persons' anomalous states of knowledge, of which the representation is an important aspect of information retrieval research. They constructed representations of information needs that stem from a description of real users' needs in the context of a literature search. These representations were then assessed by study participants.

Kuhlthau~\cite{DBLP:journals/jasis/Kuhlthau91} describes six stages of information seeking processes:
initiation, selection, exploration, formulation, collection and presentation.
Similarly, Ellis et al.~\cite{ellis} also defined several categories of information behaviour: starting, chaining, browsing, differentiating, monitoring, extracting, verifying, and ending.
Bates~\cite{Bates1989TheDO} describes information search strategies which, amongst others, contain backwards chasing, forward chasing, and the identification of central journals for areas.
Wilson~\cite{DBLP:journals/ipm/Wilson97} describes information seeking behaviour via four stages: passive attention, passive search, active search, and ongoing search.
%
Weigl et al.~\cite{DBLP:conf/jcdl/WeiglPOD17} describe the correspondence and gaps between the models of Ellis et al.~\cite{ellis}, Bates~\cite{Bates1989TheDO} and Wilson~\cite{DBLP:journals/ipm/Wilson97}.

Vakkari~\cite{Vakkari} describes the formulation of information needs as an iterative process. Users acquire new information, influencing their perception of the information space.
He states that the evaluation of exploratory search systems should focus on the impact they have on users during this information search process and how the retrieved information furthers the users' task. He introduces measures to quantify the change in a user's search strategy between query formulations or sessions. 

Taylor~\cite{DBLP:journals/ipm/Taylor12} examines which different factors users of information search systems consider when making relevancy decisions in an information gathering process and when which factors are relevant.
The search stages observed are the ones described by Ellis~\cite{ellis} and Kuhlthau~\cite{kuhlthau2004seeking}.
Study participants conducted individual web searches where they chose their search stage, the relevancy of results, and the relevancy criterion that most influenced their decision out of 19 predefined ones following previous literature~\cite{Cool1993CharacteristicsOT,DBLP:journals/ipm/BarryS98}.

More recent research considers information seeking by users as a dynamic process where users' goals and intentions vary depending on their current search or exploration steps~\cite{DBLP:conf/sigir/Azzopardi14,DBLP:conf/jcdl/LiuS22,DBLP:journals/tois/RuotsaloPEGFMJK18}.

Contrasting these works, we propose comparing different expression levels of complex information needs with multiple relevancy indicators. This could highlight and capture changes between these levels of the same information needs instead of identifying stages and indicators for many random information search tasks.

\section{Concept}

We propose observing different stages of the expression of users' typical information needs in digital libraries, such as \textit{"what papers are about or fit a specific topic"}~\cite{DBLP:conf/jcdl/KreutzBS22,DBLP:conf/acl/BettsPA19,DBLP:conf/ercimdl/BloehdornCDHHTV07,zhu}.
Contrasting recent work~\cite{DBLP:conf/sigir/Azzopardi14,DBLP:conf/jcdl/LiuS22,DBLP:journals/tois/RuotsaloPEGFMJK18} our concept does not only focus on queries posed to DL interfaces. We explicitly assume that the actual visible interaction with a system is only part of the complete information seeking process~\cite{ingwersen}. An initial query was preceded by some internal considerations or general tendencies of users.

\subsection{Manifestations}
Using Taylor's~\cite{Taylor1968QuestionNegotiationAI} four-level continuum for simplicity, we try to map out the levels to capture:
\textit{First}, we propose to observe the personal \textbf{definition} of an information need without the satisfaction of the information need in mind, e.g., \textit{"define relevancy of a paper for a topic"}. This is a conscious verbalisation of important factors that a user considers relevant when generally thinking about a specific information need. We expect this manifestation to lie between levels $Q_2$ and $Q_3$ as this is a notion of the general requirements, which data is required to satisfy an information need. However, there still is no verbalisation of the task conduction itself.

\textit{Second}, we propose observing the \textbf{ideal} or general satisfaction of an information need without the restriction of the scope of one specific information system, e.g., \textit{"describe your general process of finding relevant papers from a topic of your choice"}.
We regard this as a manifestation of a point between levels $Q_3$ and $Q_4$ as there is a conscious verbalisation, but no restrictions are imposed which stem from the specialisation of using one information system to satisfy the information need. A person describing their ideal task conduction might not necessarily verbalise the considered factors for their relevance decision.

\textit{Third}, we propose observing the \textbf{actual} satisfaction of an information need with the restrictions of one specific information system, e.g., \textit{"use this system to find relevant papers from a topic of your choice"}. We estimate this manifestation to correspond to level $Q_4$. In the task conduction, we assume persons unconsciously using or mentioning factors that they consider to determine the relevancy of information objects.

Additionally, as a \textit{fourth} manifestation we propose a time-delayed \textbf{re-definition} of the general information need, i.e., a second iteration of the first manifestation, could also be observed. This enables analysis of a temporal dimension.

\subsection{Data Capture and Preparation}
To capture these viewpoints, semi-structured interviews can be applied for the definition, ideal task satisfaction and re-definition manifestation. For the actual task satisfaction, a think-aloud interview with screen capture while the user conducts the task with the respective information system can be applied.
With this information, formal process models~\cite{weyers} can be derived for the ideal and actual task satisfaction processes.

From the interviews and formal task models, fine-grained factors, e.g., \textit{a paper's citations from other highly cited papers} or \textit{a paper's citations from other papers' introductions}, can be derived. To abstract from these detailed descriptions and find underlying general motives, the factors can be grouped into categories, e.g., \textit{a paper's citations}.

\subsection{Data Analysis}

With these viewpoints the differences and similarities between manifestations and, therefore, the stability of features over levels of expressions of information needs can be assessed. Underlying motives which are present throughout all scenarios or which are stable over time could be identified. Identifying and understanding these tendencies, in turn, would help in the construction of digital library systems. E.g., if users are faced with the task of determining the relevancy of a paper for a topic, they could, in general, consider changeless factors (such as \textit{title} and \textit{abstract} of a paper), those which develop over time (such as the \textit{citation count} or the \textit{perception in social media}) or ones depending on authors of these relevant papers (such as their \textit{co-authors} or \textit{affiliations}). Knowledge of the general importance of groups of factors or single factors in decision making processes of users could help shape interface designs or enable more realistic user simulations.


%

\section{Conclusion}

In this work, we propose capturing the stability of information needs by comparing the multiple stages of users' expressions: the general definition of an information need, users' ideal task satisfaction, users actually conducting a task and a re-definition of users' perception of the information need.

As a next step, we intend to capture the described manifestations via user studies to investigate common underlying motives in fulfilling information needs using digital libraries.

\bibliographystyle{IEEEtran}


\end{document}